\documentclass[aps,prd,preprint,showpacs,showkeys,preprintnumbers,amsmath,amssymb]{revtex4-1}

\usepackage{amsmath}
\usepackage{amssymb}
\usepackage{graphicx}
\usepackage{url}
\usepackage{xcolor}

\newcommand{\bef}{\begin{figure}}
\newcommand{\eef}{\end{figure}}
\newcommand{\bec}{\begin{center}}
\newcommand{\eec}{\end{center}}
\newcommand{\be}{\begin{equation}}
\newcommand{\ee}{\end{equation}}

\begin{document}

\title{On Quasinormal Modes for Scalar Perturbations of Static Spherically Symmetric Black Holes
in Nash Embedding Framework}
\author{Sergio C. Ulhoa }
\email{sc.ulhoa@gmail.com} \affiliation{Instituto de F\'{i}sica, Universidade de Bras\'{i}lia, 70910-900, Bras\'{i}lia, DF, Brazil.}

\author{Ronni G. G. Amorim }
\email{ronniamorim@gmail.com} \affiliation{Faculdade Gama, Universidade de Bras\'{i}lia, 72444-240, Setor Leste (Gama), Bras\'{i}lia, DF, Brazil.}

\author{Abra\~{a}o J. S. Capistrano }
\email{abraao.capistrano@unila.edu.br} \affiliation{Federal University of Latin-American Integration, \\
Casimiro Montenegro Filho Astronomy Center, Itaipu Technological Park, 85867-970, P.o.b: 2123, Foz do Iguassu, PR, Foz do Igua\c{c}u-PR, Brazil}

\begin{abstract}
In this paper we investigate scalar perturbations of black holes embedded in a five dimensional bulk space. It is calculated the quasinormal frequencies of a such black holes using the third order of Wentzel, Kramers, Brillouin (WKB) approximation for scalar perturbations. The results are presented in tables along the text.
\end{abstract}


\maketitle
\section{Introduction}
\noindent

The black hole solutions can be perturbed by some matter field, then it leads to oscillations, which can be explained by an analogy with a vibrating string, in such a system the oscillations decrease with time as the string loses energy to the environment \cite{berti}. Similarly to the string, where its frequency can be expressed by a complex parameter called quasi-normal frequency, the perturbations of black holes evolve with time as an oscillation with complex frequency, in  which the  real part is responsible for the period of oscillation and the imaginary part is responsible for damping or amplification. Thus proper oscillations of black holes are also called quasi-normal modes.
	
Historically, the theoretical study about perturbation of black holes begins with the seminal paper of Regge and Wheeler in 1957, in which they studied the stability of Schwazschlid black hole submitted to small perturbations \cite{wheeler}. In the 1970's, the quasi-normal modes were used by Vishveshwara, in calculations of scattering of gravitational waves by a Schwazschild black hole \cite{Vish}.  Since then, the study of this issue were directed to several problems. Black holes stability is an example of this importance. The relevance of studying the stability of these astrophysical objects is the possibility to understand theories in higher dimensions as aspects of brane-world and string theory\cite{Kon, Martel} and variants. Since the uniqueness theorem   is not applicable to more than four dimensions, stability can be the criterion to select physical solutions \cite{Miranda, Kodama}. Another important application of  the quasi-normal modes of black holes is in the study of the ADS/CFT correspondence \cite{Maldacena}, whereby the imaginary part of the quasi-normal fundamental frequency describe the thermalization time in a conformal invariant field theory in the border of the anti de-Sitter space-time. Recently, the quasi-normal modes have been related to quantum gravity; some evidences show that highly damped quasi-normal modes can be important in the attempts to quantize the area of black holes \cite{quantum}. In addition to the advantages presented, there are prospects that in the near future oscillations of astrophysical black holes will be observed using gravitational waves detectors \cite{detection}. This fact is relevant in the sense that quasi-normal modes carry information about  stellar objects, such as mass, charge and angular momentum.
	
Considering the relevance of the quasi-normal modes, in this work we study scalar perturbation of a black hole embedded in a five-dimensional bulk space by calculating its quasi-normal frequencies. In this sense, this paper is organized as follows. In section 2, we use the Nash embedding theorem to study the induced four dimensional equations embedded in a five-dimensional bulk space. In section 3, we present the induced potential in a embedded spherically symmetric vacuum solution. In section 4, we discuss about scalar perturbations. In section 5, we present the quasi-normal modes for treated system using WKB approach and using this result we discuss about black hole stability. In section 6, we present our concluding remarks.

\section{Embedded induced four dimensional equations}
The recent proof of the Poincar\`{e} conjecture by G. Perelman \cite{perelman1, perelman2} suggests a new paradigm for geometry and in particular for Einstein's gravitation and cosmology, namely the possibility that we can deform space-times in arbitrary directions, changing its shape. It has it origins from a  solution  for the  Riemann  curvature  ambiguity was conjectured by  L. Schlaefli in 1871,  proposing that the  Riemann  manifolds  should  be   embedded in a larger  bulk   space,   such  that its   Riemann curvature $\! {\cal I}\!\!{\cal R}$    would  act  as  a curvature reference  for  all
embedded  manifolds,  just  like the flat  Euclidean space  $I\!\!R^3$ acts as  a curvature reference  for  surfaces. The   Schlaefli  conjecture is  the  origin  of the embedding problem for  Riemann's  geometry.  Formally   we may  write his  proposal  as
\begin{equation}\label{eq:riemann}
{\cal I}\!\!{\cal R}(U,V)W\! =\! R(U,V)W \!+\! \mbox{extrinsic terms}
\end{equation}
To  detail,  we  must   re-write  this  expression  in terms of components  in a  suitable  reference  frame. This can be  done  in an  arbitrary   vielbein   defined  in  the bulk,  and  then  separating the bulk  curvature tensor  in normal and  tangent  components. With the  purpose of understanding the meaning of  the   Schlaefli   proposal  we  write  these  components in the Gaussian  frame   defined by  the  embedding map  itself.  In this  frame  the  components of the  extrinsic  terms   appear  as the  extrinsic  curvature component. This allows us to relate different geometries with different properties though a differential embedding. In this paper, we explore the consequences of such embedding using the dynamics properties of the extrinsic curvature. In order to make an appropriate embedding between smooth (differentiable) geometries, we use the Nash embedding theorem \cite{Nash} in order to propose a new theoretical structure able to relate it to a physical theory.

Nash showed that any embedded perturbed metric $\tilde{g}_{\mu\nu}$ can be generated by a continuous sequence of small metric perturbations of a given
initially unperturbed geometry ${g}_{\mu\nu}$ by means of
\begin{equation}\label{eq:York}
\tilde{g}_{\mu\nu}  =  g_{\mu\nu}  +  \delta y^a \, k_{\mu\nu  a}  +
\delta y^a\delta y^b\, g^{\rho\sigma}
k_{\mu\rho a}k_{\nu\sigma b}+\cdots\;,
\end{equation}
or, equivalently,
\begin{equation}\label{eq:York2}
k_{\mu\nu}=-\frac{1}{2}\frac{\partial g_{\mu\nu}}{\partial y} \;,
\end{equation}
where $y$ is the coordinate related to extra dimensions. Since Nash's smooth deformations are applied to the embedding process, the coordinate $y$, usually noticed in ridig embedded models, e.g \cite{RS,RS1}, can be omitted in the process for perturbing the element line. This seems particulary interesting to astrophysical and cosmological proposes in which traditionally the gravitational perturbation mechanisms are essentially
plagued by coordinate gauges due to the group of diffeormorphisms.

The Einstein-Hilbert principle leads  to D-dimensional Einstein's  equations  for the  bulk
metric  ${\cal G}_{AB}$  in  arbitrary  coordinates
 \be
{\mathcal{R}}_{AB}-\frac{1}{2}\mathcal{R} \mathcal{G}_{AB}
 =G_* T^*_{AB}\;,\label{eq:EEbulk}
\ee
where  we  have dispensed the bulk  cosmological constant and $T^*_{AB}$ denotes the energy-momentum tensor of the known  matter and gauge fields. The  constant  $G_*$ determines the D-dimensional energy scale. For the present application, capital Latin indices run from 1 to 5. Small case Latin indices refer to the only one extra dimension considered. All Greek indices refer to the embedded space-time counting from 1 to 4.

Concerning the confinement, the  four-dimensionality of the  space-time  is an experimentally  established fact associated  with  the   Poincar\'{e}
invariance of  Maxwell's  equations and  their  dualities, also valid for Yang-Mills gauge fields restricted to four-dimensions \cite{donaldson,taubes}. Even though the duality properties can be mathematically extended to higher dimensions, we adopt it as a condition based on experimental backgrounds \cite{lim}. Therefore, all matter, which interacts   with these gauge  fields, must  for  consistency  be also   defined in  the  four-dimensional  space-times. This consideration complements a physical interpretation for the Nash theorem that provides an interesting mechanism for perturbing and creating new geometries. In this five-dimensional framework present here, only the components that access higher dimensions are related to the extrinsic curvature $k_{\mu\nu}$, while the metric components are confined to the geometry itself, as shown in Eq.(\ref{eq:York}).  On the other hand, as well known, in  spite of  all  efforts  made so far, the gravitational interaction  has failed to  fit into  a  similar  gauge  scheme,  so that the  gravitational  field  does  not necessarily have the same
four-dimensional  limitations,  regardless  the  location of  its  sources.

In order  to   recover  Einstein's gravity  by reversing the  embedding, the   confinement  of  ordinary matter and  gauge  fields implies that the  tangent  components of  $G_* T^*_{AB}$  in the  above
equations must  coincide  with  $(8\pi G  T_{\mu\nu})$ where
$T_{\mu\nu}$  is the   energy-momentum  tensor of the confined
sources. As it may   have  been already noted,   we  are
essentially  reproducing a framework similar to brane-world program \cite{add},  with the
difference that we apply a dynamical differential embedding and  has  nothing to  do with branes as those defined in string/M theory. Using the Nash embedding theorem  together  with  the  four-dimensionality of gauge fields, one can obtain the  Einstein-Hilbert principle for the bulk and a D-dimensional energy scale $G_*$.

In addition, one can define a five-dimensional local embedding with an embedding map $\mathcal{Z}:V_{4}\rightarrow V_{5}$. We admit that $\mathcal{Z}^{\mu}$ is a regular and differentiable map with $V_{4}$ and $V_5$ being the four embedded space-time and the bulk, respectively. The components $\mathcal{Z}^{A} =\;f^{A}(x^{1},...,x^{4})$ associate with each point of $V_{4}$ a point in $V_{5}$ with coordinates $\mathcal{Z}^{A}$. These coordinates are the components of the tangent vectors of $V_{4}$. Accordingly, calculating the components of Eq.(\ref{eq:EEbulk}) one can find the induces equations for the embedded  geometry
 \begin{eqnarray}
&&R_{\mu\nu}-\frac{1}{2}R g_{\mu\nu}-
{Q}_{\mu\nu} = -8\pi G T_{\mu\nu} \label{eq:BE1}\\
 &&k_{\mu ;\rho}^{\rho}\!
 -\!h_{,\mu} = 0\label{eq:BE2}\;,
\end{eqnarray}
where now $T_{\mu\nu}$ is the  energy-momentum tensor of the confined matter. The quantities  $h^2=h.h$ with $h= g^{\mu\nu}k_{\mu\nu}$ and $K^{2}=k^{\mu\nu}k_{\mu\nu}$ are the mean curvature and Gaussian curvature, respectively. Moreover, one defines
\begin{equation}
Q_{\mu\nu} = k^{\rho}{}_{\mu }k_{\rho\nu }-h
k_{\mu\nu}\!\!-\!\!\frac{1}{2}(K^{2}-h^{2})g_{\mu\nu}\label{eq:Qij}.
\end{equation}
This  tensor  is  independently conserved, as it can be  directly verified that
(semicolon  denoting  covariant  derivative  with respect  to  $g_{\mu\nu}$)
\begin{equation}\label{eq:qmunu}
Q^{\mu\nu}{}_{;\nu} =0\;.
\end{equation}
A detailed derivation of these equations can be found in \cite{GDE,gde2,QBW} and references therein as well the higher dimensional case. Hereafter, we use a system of unit such that $c=G=1$.

\section{Induced potential in a spherically symmetric vacuum solution}
As shown in a previous work \cite{cap2014}, we start with the general static spherically symmetric induced metric that can be described by the line element as
\begin{equation}\label{eq:line element}
ds^2 = B(r) dt^2 -  A(r) dr^2 - r^2 d\theta^2 - r^2 \sin^2\theta d\phi^2\;\;,
\end{equation}
where we denote the functions $A(r)=A$ and $B(r)=B$. Thus, one can obtain the following components for the Ricci tensor:
$$R_{rr} = \frac{B''}{2B} - \frac{1}{4}\frac{B'}{B} \left(\frac{A'}{A} + \frac{B'}{B} \right) - \frac{1}{r} \frac{A'}{A}$$
$$R_{\theta\theta} = -1 + \frac{r}{2A} \left(-\frac{A'}{A} + \frac{B'}{B} \right) + \frac{1}{A}$$
$$R_{\phi\phi} = \sin^2\theta R_{\theta\theta}$$
$$R_{tt} = -\frac{B''}{2A} + \frac{1}{4}\frac{B'}{A} \left(\frac{A'}{A} + \frac{B'}{B} \right) - \frac{1}{r} \frac{B'}{A}$$
where we have $\frac{dA}{dr} = A'$ e $\frac{dB}{dr} = B'$.

From Eq.(\ref{eq:BE1}), the gravitational-tensor vacuum equations (with $T_{\mu\nu}=0$) can be written in alternative form as
\begin{equation}\label{eq:1}
    R_{\mu\nu} + \frac{1}{2} Q g_{\mu\nu} = Q_{\mu\nu}
\end{equation}
where we use the contraction $Q= g^{\mu\nu} Q_{\mu\nu}$.

The general solution of Codazzi equations Eq.(\ref{eq:BE2}) is given by
\begin{equation}\label{eq:geneqk}
    k_{\mu\nu}=f_{\mu}g_{\mu\nu} \;\;\;\;(no\;sum\;on\;\mu)\;,
\end{equation}
Taking the former equation and the definition of $Q_{\mu\nu}$, one can write
$$Q_{\mu\nu}= f^2_{\mu}g_{\mu\nu}-\left(
\sum_{\alpha}f_{\alpha}\right)f_{\mu}g_{\mu\nu}-\frac{1}{2}\left(\sum_{\alpha}f^2_{\alpha}
-\left(\sum_{\alpha}f_{\alpha}\right)^2\right)g_{\mu\nu}\;,$$
where
$$U_{\mu}=f^2_{\mu}-\left( \sum_{\alpha}f_{\alpha}\right)f_{\mu}
-\frac{1}{2}\left(\sum_{\alpha}f^2_{\alpha}-\left(\sum_{\alpha}f_{\alpha}\right)^2\right)\delta^\mu_\mu\;.$$
Consequently, we can write $Q_{\mu\nu}$ in terms of $f_{\mu}$ as
\begin{equation}
Q_{\mu\nu}= U_{\mu}g_{\mu\nu} \;\;\;\;\;(no\;\;sum\;\;on\;\;\mu).
\end{equation}

A straightforward consequence of the homogeneity of Codazzi equations Eq.(\ref{eq:BE2}) in five-dimensions is that the individual arbitrariness of the functions $f_{\mu}$ can be reduced to a unique arbitrary function $\alpha$ that depends on the radial coordinate. Hence, the equation (\ref{eq:geneqk}) turns to be
\begin{equation}\label{eq:alphak}
 k_{\mu\nu} = \alpha(r) g_{\mu\nu}\;,
\end{equation}
with $\alpha(r)=\alpha$. With a straightforward calculation of the former equations, one can obtain the coefficients of the metric as
\begin{equation}\label{eq:br1}
B(r) = 1 + \frac{K}{r} + \frac{9}{r} \int \alpha^2(r) r^2 dr\;,\\
\end{equation}
and
\begin{equation}\label{eq: ar}
A(r)= \left[B(r)\right]^{-1} = \left[ 1 + \frac{k}{r} + \frac{9}{r} \int \alpha^2(r) r^2 dr\; \right]^{-1}\;,
\end{equation}
where $k$ is a constant. In order to constrain this arbitrariness, we look at the characteristics of the extrinsic curvature itself in the asymptotic limit, which will be important to an astrophysical application. The extrinsic curvature at infinity goes to a flat space obeying the asymptotically conformal flat condition. This can be understood as the following form
\begin{equation}\label{eq:flatcondition}
\lim_{r\rightarrow \infty} k_{\mu\nu} = \lim_{r\rightarrow \infty} \alpha(r)\lim_{r\rightarrow \infty} g_{\mu\nu}\;.
\end{equation}
Since the function $\alpha(r)$ must be analytical at infinity, one can write the simplest option
\begin{equation}\label{eq:alfaform}
\alpha(r)= \sum_{n=i}^{s}\frac{\sqrt{-\alpha_0}}{\gamma^{*} r^n}\;,
\end{equation}
where the sum is upon all scalar potentials and the indices $i$ and $s$ are real numbers. Since these scalar potentials have its origin in the extrinsic curvature they do not remain confined in the embedded geometry propagating in the extra-dimension. The index $n$ represents all the set of scalar fields that fall off with $r$-coordinate following the inverse $n^{th}$ power law. The equation (\ref{eq:alfaform}) is essentially the representation of the effect of extrinsic curvature leading to a local modification of the space-time without producing umbilical point as expected for a spherical geometry. Depending on the variation of the function $\alpha(r)$ one can have a bent or stretched geometry without ripping off the manifold, and, curiously, in the same notion as pointed by Riemann himself \cite{riemann}. It was shown that the parameter $\alpha_0$ has cosmological magnitude \cite{cap2014,cap1}, i.e., it does not depend on individual astrophysical properties and has the same units as the Hubble constant. Its modulus is of the order of $0.677\;km.s^{-1}.Mpc^{-1}$. In addition, to keep the right dimension of Eq.(\ref{eq:alfaform}) we have  introduced a unitary parameter $\gamma^{*}$ that has the inverse unit of Hubble constant and also establishes the cosmological horizon in Eq.(\ref{eq:line element}). It is important to stress that the form of the function $\alpha(r)$ could be no other since the relativistic effects in astrophysical scale is estimated to be $10^{-8}$ times weaker than Newtonian ones \cite{yamada} and it must obey a distance decaying law $(\sim 1/r)$ or their smooth deviations $(\sim 1/r^2 , \sim 1/r^3)$.

Using Eqs.(\ref{eq:br1}) and (\ref{eq:alfaform}), one can obtain an explicit form of the coefficient $B(r)$  given by
\begin{equation}\label{eq:br2}
B(r) = 1 + \frac{k(9\alpha^2_0 + 1)}{r} - \sum_{n=i}^{s}\frac{9\alpha_0}{\gamma^{*}(2n-3)} r^{2\left(1-n\right)}\;.\\
\end{equation}
In terms on the correspondence principle with Einstein gravity, we set $k(9\alpha^2_0 + 1)=-2M$, which remains valid even in the limit when $\alpha_0\rightarrow 0$ in order to obtain the asymptotically flat solution.

\section{Scalar Perturbations}\label{GP}
\noindent
In this section, we focus our attention on the study of Eq.(\ref{eq:br2}) under scalar perturbations. The study of quasinormal modes can be treated by scalar fields and appropriate metrics. In this section, we construct the Klein-Gordon equation in a curved space-time. In order to achieve such a goal, we consider the line element in a form
\begin{equation}\label{metric}
ds^2=f(r)dt^2-f(r)^{-1}dr^2+r^2d\Omega.
\end{equation}
It is well known that the massless Klein-Gordon equation is given by
\begin{equation}\label{kg1}
\nabla_{\mu}\nabla^{\mu}\Phi=0,
\end{equation}
or, equivalently
\begin{equation}\label{kg2}
\frac{1}{\sqrt{-g}}\partial_{\mu}(\sqrt{-g}g^{\mu\nu}\partial_{\mu}\Phi)=0,
\end{equation}
where $\Phi=\Phi(r,\theta, \phi, t)$.
Using Eq.(\ref{metric}), we can write
\begin{equation}\label{metric2}
\sqrt{-g}=r^2\sin\theta.
\end{equation}
Moreover, Eq.(\ref{kg2}) can be written as
\begin{equation}\label{kg3}
\frac{1}{f(r)}\frac{\partial^2\Phi}{\partial t^2}-\frac{\partial}{\partial r}\left(f(r)\frac{\partial\Phi}{\partial r}\right)- \frac{2f(r)}{r}\frac{\partial\Phi}{\partial r}- \frac{1}{r^2\sin\theta}\frac{\partial}{\partial\theta}\left(\sin\theta\frac{\partial\Phi}{\partial\phi}\right)-\frac{1}{r^2\sin^{2}\theta}\frac{\partial^{2}\Phi}{\partial\phi^2}=0.
\end{equation}
Taking the ansatz
\begin{equation}\label{ansatz}
\Phi(r,\theta,\phi, t)=\sum_{l=0}^{\infty}\sum_{m=-l}^{l}Y_{l,m}(\theta,\phi)e^{-i\omega t}\frac{\psi_{l,m}(r)}{r},
\end{equation}
we obtain
\begin{equation}\label{kg4a}
\left[\frac{1}{\sin\theta}\frac{\partial}{\partial\theta}\left(\sin\theta\frac{\partial\Phi}{\partial\phi}\right)-\frac{1}{\sin^{2}\theta}\frac{\partial^{2}\Phi}{\partial\phi^2}\right]Y_{l,m}(\theta,\phi)=-l(l+1)Y_{l,m}(\theta,\phi),
\end{equation}
where $Y_{l,m}(\theta,\phi)$ are the spherical harmonics.
\begin{equation}\label{kg4b}
\left(\frac{-\omega^2}{f(r)}+\frac{l(l+1)}{r^2}-f(r)\frac{\partial^2}{\partial r^2}-f'(r)\frac{\partial}{\partial r}-\frac{2f(r)}{r}\frac{\partial}{\partial r}\right)\frac{\psi_{l,m}(r)}{r}=0.
\end{equation}
Taking the following coordinates change  $dx=dr/f(r)$, we obtain
\begin{equation}\label{kg5}
\frac{d^2\psi_{l,m}(r)}{dx^2}+(\omega^2-V(x))\psi_{l,m}(r)=0,
\end{equation}
where we denote $$V(x)=f(x)\frac{l(l+1)}{x^2}+f(x)\frac{f'(x)}{x}.$$

If we consider equation (\ref{eq:br2}), then we finally get
\begin{equation}\label{f}
f(r)=1-\frac{2M}{r}+\sum_{j}\frac{\alpha_{0}}{2j-3}r^{2(1-j)}.
\end{equation}

\begin{figure}[!htb]
\includegraphics[scale=0.5]{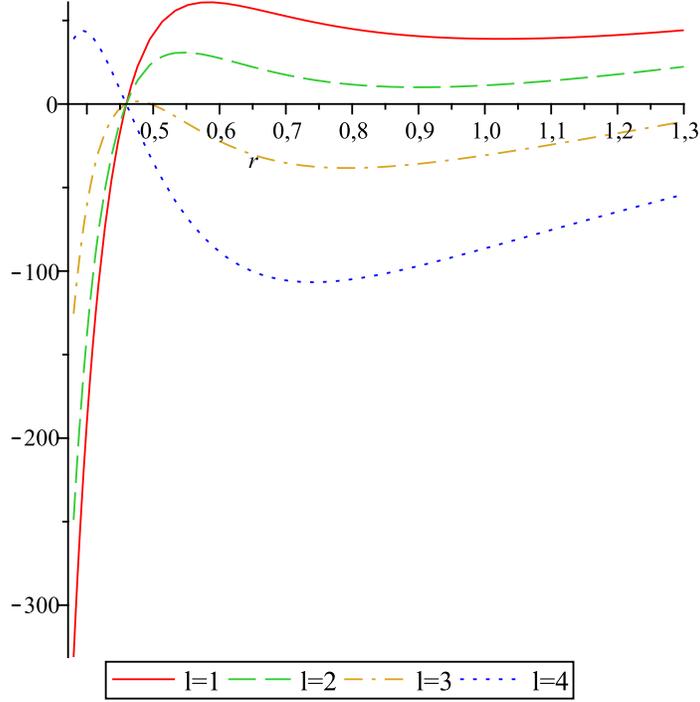}
\caption{The potential $V$ in function of the radius for values $l=1$ (solid thick line), $l=2$ (long-dashed line), $l=3$ (short-dashed line) and $l=4$ (solid thin line) for the fixed values of $\alpha=0.677$, $M=1$ and $\mu=0$.}
\label{fig:fig01}
\end{figure}

\begin{figure}[!htb]
\includegraphics[scale=0.5]{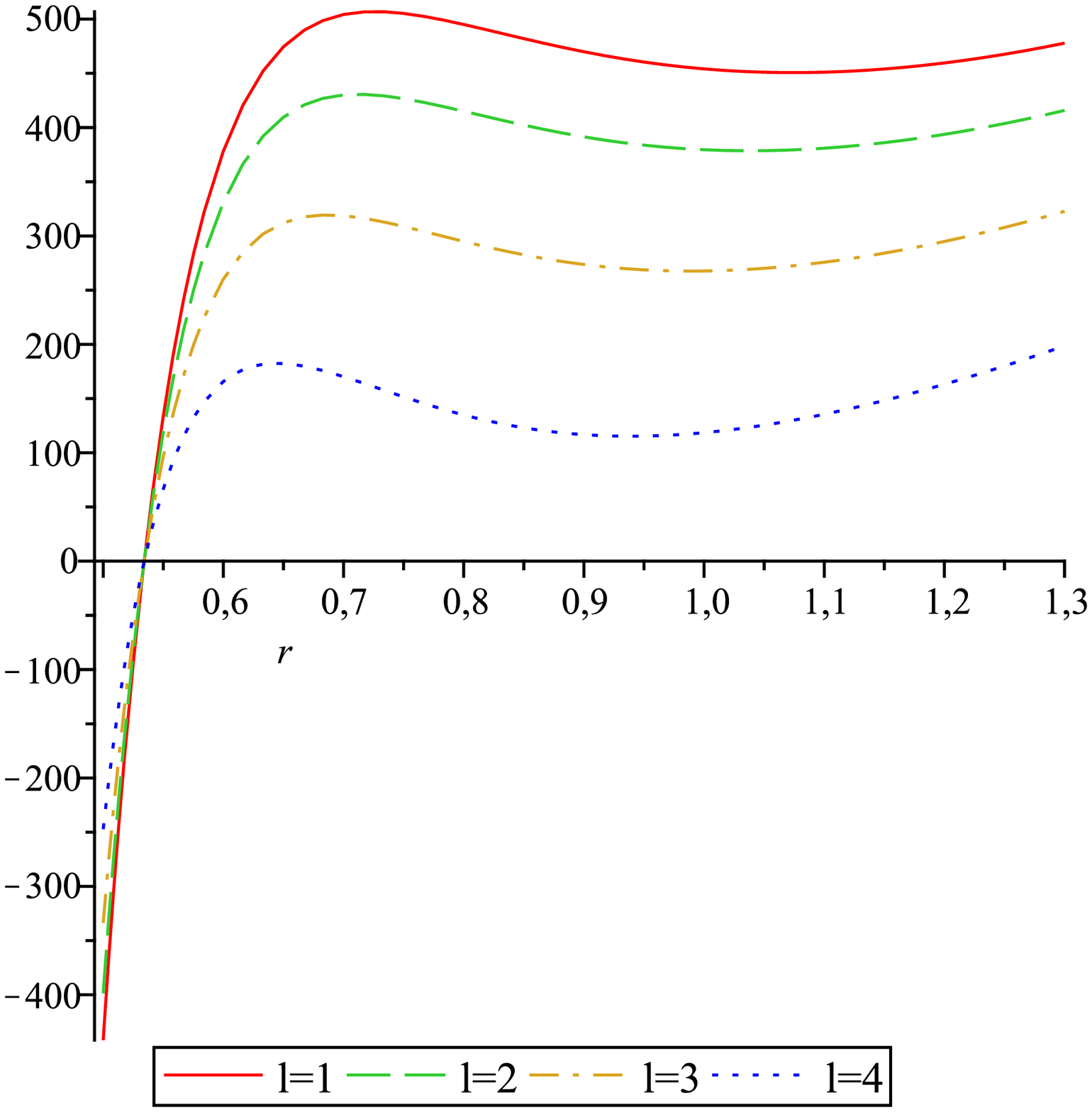}
\caption{The potential $V$ in function of the radius for values $l=1$ (solid thick line), $l=2$ (long-dashed line), $l=3$ (short-dashed line) and $l=4$ (solid thin line) for the fixed values of $\alpha=2$, $M=1$ and $\mu=0$.}
\label{fig:fig2}
\end{figure}

Thus once the potential $V$ is well defined, we can plot its behaviour for different values of $\alpha$ and $l$, which are given in fig.(\ref{fig:fig01}) and fig.(\ref{fig:fig2}).

\begin{figure}[!htb]
\includegraphics[scale=0.5]{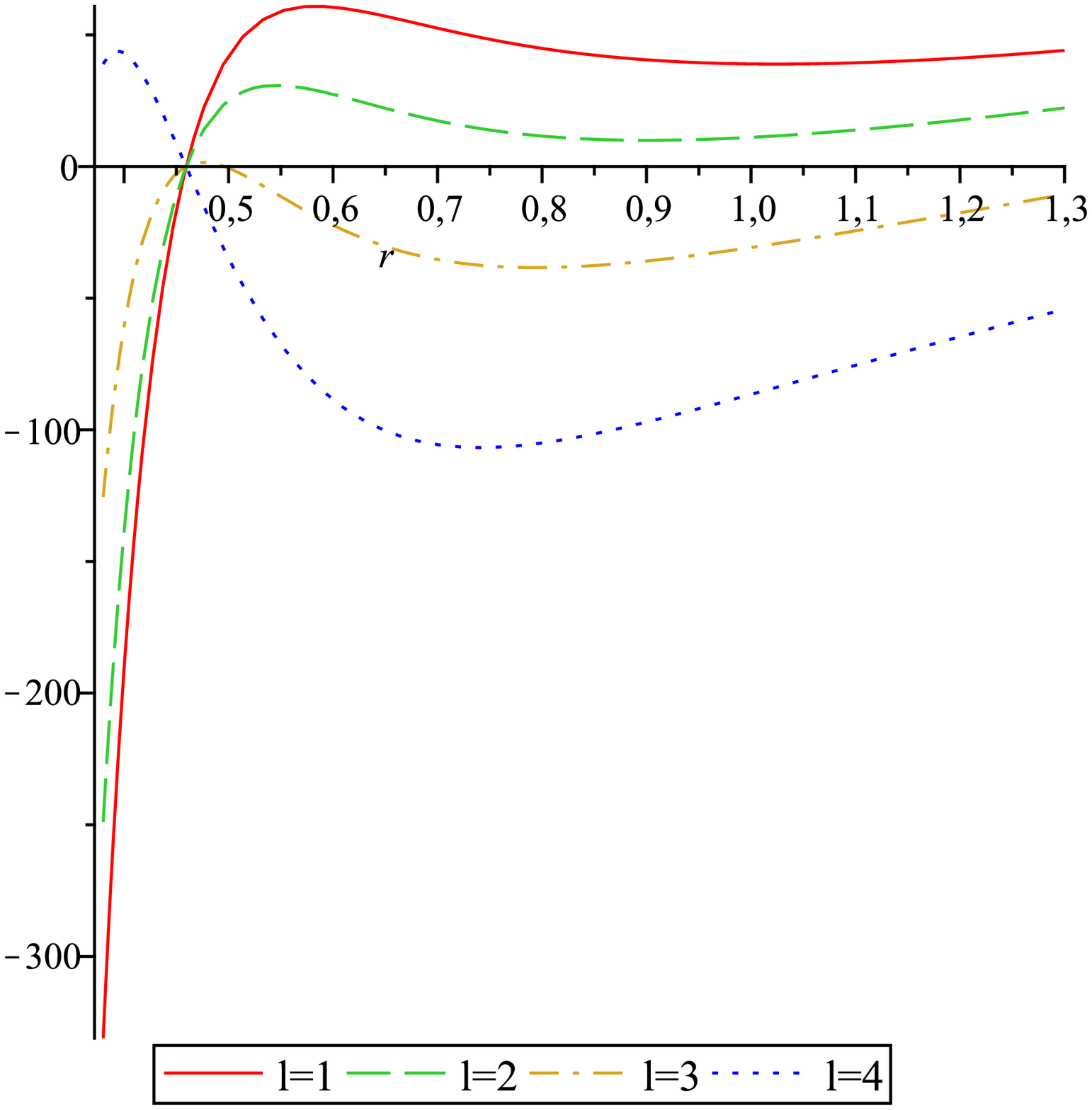}
\caption{The potential $V$ in function of the radius for values $l=1$ (solid thick line), $l=2$ (long-dashed line), $l=3$ (short-dashed line) and $l=4$ (solid thin line) for the fixed values of $\alpha=0.677$, $M=1$ and $\mu=0.1M$.}
\label{fig3}
\end{figure}

For a massive field the potential is modified to $V(r)\rightarrow V(r)+\mu^2f(r)/r$, the angular part remains the same. In fig.(\ref{fig3}) we plot such a potential for $\alpha=0.677$, $M=1$ and $\mu=0.1M$.

\section{Quasinormal Modes in WKB Approximation}\label{QNM}
\noindent

\begin{table*}
\caption{ Quasinormal modes of scalar perturbations for $\alpha=0.677$\, with $M=1$.}
        \begin{tabular}{ccccccccccc}
    \hline\hline
    $l$ & $$ & $n$ & $$ & $\omega_{nl}$& $$ & $l$ & $$ & $n$ & $$ & $\omega_{nl}$ \\
    \hline
    $1$&$$&$0$&$$&$4.444548318-8.863420102i$&$$&$4$&$$&$0$&$$&$2.789319410+26.91329506i$\\
    $$&$$&$1$&$$&$11.59600413-26.25141968i$&$$&$$&$$&$1$&$$&$136.9620887+165.7373689i$\\
    $2$&$$&$0$&$$&$2.556070877-9.893346520i$&$$&$$&$$&$2$&$$&$317.9204012+354.6421808i$\\
    $$&$$&$1$&$$&$6.342841226-24.95915206i$&$$&$$&$$&$3$&$$&$540.1554148+583.4781202i$\\
    $$&$$&$2$&$$&$12.94011467-41.62859430i$&$$&$5$&$$&$0$&$$&$6.400241451+55.35392070i$\\
    $3$&$$&$0$&$$&$0.3142845589+12.96051935i$&$$&$$&$$&$1$&$$&$301.0022385+362.7954640i$\\
    $$&$$&$1$&$$&$38.84486810-53.32057992i$&$$&$$&$$&$2$&$$&$697.9325995+777.2504936i$\\
    $$&$$&$2$&$$&$93.40199364-111.3133472i$&$$&$$&$$&$3$&$$&$1185.464821+1279.187469i$\\
    $$&$$&$3$&$$&$160.5504266-181.4993689i$&$$&$$&$$&$4$&$$&$1749.721283+1855.936081i$\\
    \hline\hline
    \label{tabela0}
  \end{tabular}
\end{table*}

\begin{table*}
\caption{ Quasinormal modes of scalar perturbations for $\alpha=2$\, with $M=1$.}
        \begin{tabular}{ccccccccccc}
    \hline\hline
    $l$ & $$ & $n$ & $$ & $\omega_{nl}$& $$ & $l$ & $$ & $n$ & $$ & $\omega_{nl}$ \\
    \hline
    $1$&$$&$0$&$$&$28.98497167-36.65093824i$&$$&$4$&$$&$0$&$$&$14.72079936-30.28374333i$\\
    $$&$$&$1$&$$&$103.8128126-126.3594331i$&$$&$$&$$&$1$&$$&$47.53015516-80.58865151i$\\
    $2$&$$&$0$&$$&$29.20727668-38.11077359i$&$$&$$&$$&$2$&$$&$100.4540639-146.2752812i$\\
    $$&$$&$1$&$$&$106.1438323-128.6137680i$&$$&$$&$$&$3$&$$&$169.1566913-224.7764688i$\\
    $$&$$&$2$&$$&$224.2314700-254.7665296i$&$$&$5$&$$&$0$&$$&$.5248454700+28.40271014i$\\
    $3$&$$&$0$&$$&$26.35460874-37.16001393i$&$$&$$&$$&$1$&$$&$45.08904231-85.15794051i$\\
    $$&$$&$1$&$$&$95.10645314-118.6073793i$&$$&$$&$$&$2$&$$&$116.4495623-166.1450658i$\\
    $$&$$&$2$&$$&$200.6518286-232.2842502i$&$$&$$&$$&$3$&$$&$205.7296676-263.8287746i$\\
    $$&$$&$3$&$$&$333.1889473-371.1040372i$&$$&$$&$$&$4$&$$&$309.9693513-375.5150908i$\\
    \hline\hline
    \label{tabela1}
  \end{tabular}
\end{table*}

\begin{table*}
\caption{ Quasinormal modes of massive scalar perturbations for $\alpha=0.677$ and $\mu=0.1M$\, with $M=1$.}
        \begin{tabular}{ccccccccccc}
    \hline\hline
    $l$ & $$ & $n$ & $$ & $\omega_{nl}$& $$ & $l$ & $$ & $n$ & $$ & $\omega_{nl}$ \\
    \hline
    $1$&$$&$0$&$$&$29.00936098-36.67332200i$&$$&$4$&$$&$0$&$$&$14.74857813-30.30037176i$\\
    $$&$$&$1$&$$&$103.9107237-126.4501766i$&$$&$$&$$&$1$&$$&$47.63840864-80.66160380i$\\
    $2$&$$&$0$&$$&$29.23683246-38.13734008i$&$$&$$&$$&$2$&$$&$100.6803470-146.4444515i$\\
    $$&$$&$1$&$$&$106.2610002-128.7222323i$&$$&$$&$$&$3$&$$&$169.5250565-225.0711954i $\\
    $$&$$&$2$&$$&$224.4751039-254.9976125i$&$$&$5$&$$&$0$&$$&$0.5217309759+28.39953701i$\\
    $3$&$$&$0$&$$&$26.38849620-37.18864399i$&$$&$$&$$&$1$&$$&$45.02864470-85.11181068i$\\
    $$&$$&$1$&$$&$95.24043132-118.7279173i$&$$&$$&$$&$2$&$$&$116.3098696-166.0264653i$\\
    $$&$$&$2$&$$&$200.9301766-232.5433005i$&$$&$$&$$&$3$&$$&$205.4963229-263.6210918i$\\
    $$&$$&$3$&$$&$333.6438082-371.5352605i$&$$&$$&$$&$4$&$$&$309.6292522-375.2043911i$\\
    \hline\hline
    \label{tabela2}
  \end{tabular}
\end{table*}

In this section, we use the  Wentzel, Kramers, Brillouin (WKB) approximation in such a regime the frequency is given by~\cite{Konoplya:2011qq}
$$\frac{\imath\,\left(\omega^2-V_0\right)}{\sqrt{-2V_0^{\prime\prime}}}-\sum_{i=2}^{k}\Lambda_i=n+\frac{1}{2}\,,$$
where $V_0$ and $V_0^{\prime\prime}$ are the effective potential and its second derivative respectively, taken at the point of the maximum of $V$. The sum above represents high order correction in the usual WKB method~\cite{PhysRevD.68.024018}. Thus the higher is the orders of approximation the better is the result, which could be comparable to numerical methods~\cite{Konoplya:2011qq}. In this sense it is enough to work with WKB approximation of third order~\cite{PhysRevD.35.3632}, which is given by

\begin{equation}
\omega_{n,l}^2=\left[V_0+(-2V_0^{\prime\prime})^{1/2}\Lambda\right]- \imath\left(n+\frac{1}{2}\right)(-2V_0^{\prime\prime})^{1/2}(1+\Omega)\,,\label{omega}
\end{equation}
where

\begin{widetext}
\begin{equation*}
\Lambda=\frac{1}{(-2V_0^{\prime\prime})^{1/2}}\left[\frac{1}{8}\,\left(\frac{V^{(4)}_0}{V^{\prime\prime}_0}\right)\left(\frac{1}{4}+\beta^2\right)-
\frac{1}{288}\,\left(\frac{V^{\prime\prime\prime}_0}{V^{\prime\prime}_0}\right)^2\left(7+60\beta^2\right)\right]\,,
\end{equation*}
and

\begin{eqnarray}
\Omega&=&\frac{1}{(-2V_0^{\prime\prime})}\Biggl[\frac{5}{6912}\,\left(\frac{V^{\prime\prime\prime}_0}{V^{\prime\prime}_0}\right)^4
\left(77+188\beta^2\right)-\frac{1}{384}\,\left(\frac{(V^{\prime\prime\prime}_0)^2V^{(4)}_0}{(V^{\prime\prime}_0)^3}\right)
\left(51+100\beta^2\right)+\nonumber\\
&+&\frac{1}{2304}\,\left(\frac{V^{(4)}_0}{V^{\prime\prime}_0}\right)^2\left(67+68\beta^2\right)
+\frac{1}{288}\,\left(\frac{V^{\prime\prime\prime}_0V^{(5)}_0}{(V^{\prime\prime}_0)^2}\right)\left(19+28\beta^2\right)
-\frac{1}{288}\,\left(\frac{V^{(6)}_0}{V^{\prime\prime}_0}\right)\left(5+4\beta^2\right)\Biggr],\nonumber
\end{eqnarray}
\end{widetext}
Here $\beta=n+\frac{1}{2}$ and $V^{(n)}_0=\frac{d^n V}{dx^n}|_{x=x_0}$. The point $x_0$ is the solution of the equation  $\frac{dV}{dx}(x_0)=0$, which means a point of maximum of the effective potential. We point out that $x$ is the tortoise coordinate, already defined.

Next we chart our results on tables \ref{tabela0}, \ref{tabela1} and \ref{tabela2}. We see from \ref{tabela0} that the scalar  perturbation for a realistic value of $\alpha$ is not stable. This complements our previous work in which we concluded that such a solution is not globally thermodynamically stable and their stability is reduced to very constrained range\cite{bhnash}. On the other hand from tables \ref{tabela1} and \ref{tabela2} we see that the stability is achieved by a choice of a hypothetical value of $\alpha$ (in this case, $\alpha>0.667$ will lead to a more faster accelerated regime as expected for a phantom-like cosmology \cite{cap1}) or for a massive scalar field. In all scenarios the real part of the frequency increases rapidly, which indicates that the quasi-normal frequency oscillates faster for each mode. Hence, due to the positive growing of the imaginary part, such a behaviour is very similar to a resonance.


\begin{table*}
\caption{ Quasinormal modes of scalar perturbations for $\alpha=0.677$\, with $M=1$.}
        \begin{tabular}{ccccccccccc}
    \hline\hline
    $l$ & $$ & $n$ & $$ & $\omega_{nl}$& $$ & $l$ & $$ & $n$ & $$ & $\omega_{nl}$ \\
    \hline
    $1$&$$&$0$&$$&$4.444548318-8.863420102i$&$$&$4$&$$&$0$&$$&$2.789319410+26.91329506i$\\
    $$&$$&$1$&$$&$11.59600413-26.25141968i$&$$&$$&$$&$1$&$$&$136.9620887+165.7373689i$\\
    $2$&$$&$0$&$$&$2.556070877-9.893346520i$&$$&$$&$$&$2$&$$&$317.9204012+354.6421808i$\\
    $$&$$&$1$&$$&$6.342841226-24.95915206i$&$$&$$&$$&$3$&$$&$540.1554148+583.4781202i$\\
    $$&$$&$2$&$$&$12.94011467-41.62859430i$&$$&$5$&$$&$0$&$$&$6.400241451+55.35392070i$\\
    $3$&$$&$0$&$$&$0.3142845589+12.96051935i$&$$&$$&$$&$1$&$$&$301.0022385+362.7954640i$\\
    $$&$$&$1$&$$&$38.84486810-53.32057992i$&$$&$$&$$&$2$&$$&$697.9325995+777.2504936i$\\
    $$&$$&$2$&$$&$93.40199364-111.3133472i$&$$&$$&$$&$3$&$$&$1185.464821+1279.187469i$\\
    $$&$$&$3$&$$&$160.5504266-181.4993689i$&$$&$$&$$&$4$&$$&$1749.721283+1855.936081i$\\
    \hline\hline
    \label{tabela0}
  \end{tabular}
\end{table*}

\begin{table}
\caption{ Quasinormal modes of scalar perturbations for $\alpha=2$\, with $M=1$.}
        \begin{tabular}{ccccccccccc}
    \hline\hline
    $l$ & $$ & $n$ & $$ & $\omega_{nl}$& $$ & $l$ & $$ & $n$ & $$ & $\omega_{nl}$ \\
    \hline
    $1$&$$&$0$&$$&$28.98497167-36.65093824i$&$$&$4$&$$&$0$&$$&$14.72079936-30.28374333i$\\
    $$&$$&$1$&$$&$103.8128126-126.3594331i$&$$&$$&$$&$1$&$$&$47.53015516-80.58865151i$\\
    $2$&$$&$0$&$$&$29.20727668-38.11077359i$&$$&$$&$$&$2$&$$&$100.4540639-146.2752812i$\\
    $$&$$&$1$&$$&$106.1438323-128.6137680i$&$$&$$&$$&$3$&$$&$169.1566913-224.7764688i$\\
    $$&$$&$2$&$$&$224.2314700-254.7665296i$&$$&$5$&$$&$0$&$$&$.5248454700+28.40271014i$\\
    $3$&$$&$0$&$$&$26.35460874-37.16001393i$&$$&$$&$$&$1$&$$&$45.08904231-85.15794051i$\\
    $$&$$&$1$&$$&$95.10645314-118.6073793i$&$$&$$&$$&$2$&$$&$116.4495623-166.1450658i$\\
    $$&$$&$2$&$$&$200.6518286-232.2842502i$&$$&$$&$$&$3$&$$&$205.7296676-263.8287746i$\\
    $$&$$&$3$&$$&$333.1889473-371.1040372i$&$$&$$&$$&$4$&$$&$309.9693513-375.5150908i$\\
    \hline\hline
    \label{tabela1}
  \end{tabular}
\end{table}

\begin{table}
\caption{ Quasinormal modes of massive scalar perturbations for $\alpha=0.677$ and $\mu=0.1M$\, with $M=1$.}
        \begin{tabular}{ccccccccccc}
    \hline\hline
    $l$ & $$ & $n$ & $$ & $\omega_{nl}$& $$ & $l$ & $$ & $n$ & $$ & $\omega_{nl}$ \\
    \hline
    $1$&$$&$0$&$$&$29.00936098-36.67332200i$&$$&$4$&$$&$0$&$$&$14.74857813-30.30037176i$\\
    $$&$$&$1$&$$&$103.9107237-126.4501766i$&$$&$$&$$&$1$&$$&$47.63840864-80.66160380i$\\
    $2$&$$&$0$&$$&$29.23683246-38.13734008i$&$$&$$&$$&$2$&$$&$100.6803470-146.4444515i$\\
    $$&$$&$1$&$$&$106.2610002-128.7222323i$&$$&$$&$$&$3$&$$&$169.5250565-225.0711954i $\\
    $$&$$&$2$&$$&$224.4751039-254.9976125i$&$$&$5$&$$&$0$&$$&$0.5217309759+28.39953701i$\\
    $3$&$$&$0$&$$&$26.38849620-37.18864399i$&$$&$$&$$&$1$&$$&$45.02864470-85.11181068i$\\
    $$&$$&$1$&$$&$95.24043132-118.7279173i$&$$&$$&$$&$2$&$$&$116.3098696-166.0264653i$\\
    $$&$$&$2$&$$&$200.9301766-232.5433005i$&$$&$$&$$&$3$&$$&$205.4963229-263.6210918i$\\
    $$&$$&$3$&$$&$333.6438082-371.5352605i$&$$&$$&$$&$4$&$$&$309.6292522-375.2043911i$\\
    \hline\hline
    \label{tabela2}
  \end{tabular}
\end{table}

\section{Conclusion}

In this article we have analyzed the stability of the line element obtained in \cite{cap2014, bhnash}, under scalar perturbations. Such a metric was derived by searching for a solution with four-dimensional metric spherically symmetric in a five dimensional bulk space. Thus we identified an effective potential that composes the gravitational field equations. We have calculated the quasi-normal modes of Klein-Gordon equation in order to analyze the stability of our metric. We concluded that the massive scalar perturbation can yields a stable configuration or either a non realistic massless perturbation, as it is possible to see from tables \ref{tabela1} and \ref{tabela2}. We point out that the instability of the solution for massless scalar perturbation with $\alpha=0.677$ complements our previous observation about the thermodynamical features of such a solution.

\end{document}